 \documentclass[amsmath,amssymb,superscriptaddress]{nature}
 
 \usepackage{graphicx}% Include figure files
 \usepackage{dcolumn}% Align table columns on decimal point
 \usepackage{bm}% bold math
 \usepackage{color}
 \usepackage{changes}
 \usepackage{comment}
 \usepackage{cite}
 \makeatletter
 \let\saved@includegraphics\includegraphics
 \AtBeginDocument{\let\includegraphics\saved@includegraphics}
 \renewenvironment*{figure}{\@float{figure}}{\end@float}
 \makeatother
 
 \renewcommand{\figurename}{\textbf{Figure}}

 \newcommand{\tm}{C$_{8}$-DNBDT-NW}
 \newcommand{\tmfull}{3,11-dioctyldinaphtho[2,3-\textit{d}:2',3'-\textit{d}']benzo[1,2-\textit{b}:4,5-\textit{b}']dithiophene}
 \newcommand{\mobunit}{cm$^{2}$ V$^{-1}$ s$^{-1}$}
 \newcommand{\mob}{$\mu_{\mathrm{Hall}}$}
 \newcommand{\carrier}{$n_{\mathrm{Hall}}$}
 
 \newcommand{\Id}{$I_{\mathrm{D}}$}
 \newcommand{\Ig}{$I_{\mathrm{G}}$}
 \newcommand{\Vd}{$V_{\mathrm{D}}$}
 \newcommand{\Vg}{$V_{\mathrm{G}}$}

 \newcommand{\sheetc}{$\sigma_{\mathrm{sheet}}$}
 \newcommand{\sheetr}{$R_{\mathrm{sheet}}$}
 \newcommand{\QR}{$h / e^2$}
 
 \newcommand{\Vxx}{$V_{\mathrm{4T}}$}
 \newcommand{\Rxy}{$R_{\mathrm{xy}}$}
 \newcommand{\kbT}{$k_{\mathrm{B}} T$}

 \newcommand{\CP}{$\mu^{\prime}$}
 
 \title{Two-dimensional hole gas in organic semiconductors}

 \author{Naotaka Kasuya$^{1,2}$, Junto Tsurumi$^{1,3}$, Toshihiro Okamoto$^{1,2,4}$, Shun Watanabe$^{1,2}$ \& Jun Takeya$^{1,2,3}$}

\begin{document}

\maketitle

\begin{affiliations}
	\item Material Innovation Research Center (MIRC) and Department of Advanced Materials Science, Graduate School of Frontier Sciences, The University of Tokyo, 5-1-5 Kashiwanoha, Kashiwa, Chiba 277-8561, Japan 
	\item AIST-UTokyo Advanced Operando-Measurement Technology Open Innovation laboratory (OPERAND-OIL), National Institute of Advanced Industrial Science and Technology (AIST), 5-1-5 Kashiwanoha, Kashiwa, Chiba 277-8561, Japan
	\item International Center for Materials Nanoarchitectonics (WPI-MANA), National Institute for Materials Science (NIMS), 1-1 Namiki, Tsukuba, Ibaraki 305-0044, Japan
	\item Precursory Research For Embryonic Science and Technology (PRESTO), Japan Science and Technology Agency (JST), Kawaguchi, Saitama 332-0012, Japan

\end{affiliations}

\begin{abstract}
	A highly conductive metallic gas that is quantum mechanically confined at a solid-state interface is an ideal platform to explore nontrivial electronic states that are otherwise inaccessible in bulk materials ~\cite{}. Although two-dimensional electron gas (2DEG) has been realized in conventional semiconductor interfaces~\cite{}, examples of two-dimensional hole gas (2DHG), which is the counter analogue of 2DEG, are still limited~\cite{}. Here, we report the observation of a 2DHG in solution-processed organic semiconductors in conjunction with an electric double-layer using ionic liquids. A molecularly flat single crystal of high mobility organic semiconductors serves as a defect-free interface that facilitates two-dimensional confinement of high-density holes. Remarkably low sheet resistance of 6 k$\Omega$ and high hole gas density of 10$^{14}$ cm$^{-2}$ result in a metal-insulator transition at ambient pressure. The measured degenerated holes in the organic semiconductors provide a broad opportunity to tailor low-dimensional electronic states using molecularly engineered heterointerfaces.
	
\end{abstract}

\newpage

	A metallic gas confined in semiconductor heterostructures, which is also known as a two-dimensional electron gas (2DEG) or two-dimensional hole gas (2DHG), is one of the most intriguing platforms not only for exploring the fundamentals of condensed matter~\cite{ando1982electronic,Davies_2DEG_book} but also for developing high performance devices~\cite{mimura2002early,pengelly2012review}. Over the past four decades, the discovery of 2DEGs at the interface between compound semiconductors~\cite{mimura1980new,ambacher1999two} as well as at the interface between oxide insulators~\cite{ohtomo2004high,thiel2006tunable} has provided an in-depth understandings of the nontrivial electronic states that are otherwise inaccessible in a bulk material~\cite{hwang2012emergent}. Although 2DEGs with a remarkably high sheet conductivity at n-type interfaces have been discovered and are widely used in various high frequency devices (high-electron-mobility transistors are the most representative example), the use of its counter analogue, $i.e$., 2DHGs, at p-type interfaces, are still limited ~\cite{lee2018direct,chaudhuri2019polarization}. The lack of these 2DHGs is rooted in the physics of conventional semiconductor heterointerfaces; wide-gap semiconductors generally lead to heavy valence bands, resulting in low mobility holes and deep valence bands~\cite{chaudhuri2019polarization}. In addition, extra attention should be paid to lattice continuity with respect to atomic scale precision, establishment of polar discontinuity, and extreme elimination of dangling bond, whose excess electrons often suppress hole transport~\cite{lee2018direct}. \par
	
	Metallic phases under high carrier densities have been extensively studied in the field of organic semiconductors (OSCs)~\cite{chwang2001temperature,takeya2005hall,podzorov2005hall,hulea2006tunable,takeya2007crystal,alves2008metallic,xia2010carrier,wang2012hopping,sakanoue2010band}. Recent studies have shown that self-assembled molecules without any dangling bonds can construct highly periodic electrostatic potential, and a coherent band hole system is realized, even in van der Waals bonded molecular crystals~\cite{hasegawa2009organic,Tsurumi_NatPhys_2017,fratini2017map}. Despite the recent success in the field of materials science or in the developments in printing technologies for the single-crystalline forms of OSCs ~\cite{fratini2020charge}, an apparent metallic gas state has not been observed in OSCs. This is clearly because both static as well as dynamic disorders result in the carrier localization $.i.e.,$. large fluctuations in the intermolecular transfer integrals caused by thermal molecular motions are negligible even in the single crystals of OSCs. \par

	 Here, we demonstrate solution-processed, organic 2DHGs in which the metallic gas can be confined electrostatically within a monolayer of the OSCs. OSCs have been studied extensively as frontier materials for the development of new generation electronics. Recently, improvements in synthetic routes and device fabrication techniques have led to the development of small-molecule OSCs with high carrier mobility of 10 \mobunit~\cite{mitsui2014high,Tsurumi_NatPhys_2017,Yamamura_SciAdv_2018,okamoto2020bent,Kumagai_SciRep_2019,Yamamura_AdvFun_2020}. We unambiguously demonstrate a metal-insulator transition (MIT) in the solution-processed, single-crystalline OSC at ambient pressure. In this, an extremely high carrier density of approximately 1 $\times$ 10$^{14}$ cm$^{-2}$ (approaching 0.25 holes per molecule) can be confined two-dimensionally at the OSC/electric double-layer (EDL) interface. An apparent metallic signature, $i.e.,$ the positive temperature coefficient of the resistance ($dR/dT$ $>$ 0), is observed concomitantly with a minimum resistance of 6 k$\Omega$ (below the two-dimensional quantum value \QR, $e$ is the elementary charge and $h$ is the Planck constant) down to a temperature of $T$ = 15 K. This has not yet been demonstrated in a single component of OSCs. This observation is in striking contrast to the localized nature of the electronic states in OSCs and is manifested such that the degenerated electronic states are realized in OSCs. \par

\begin{figure}[htbp]
		\centering
		\includegraphics[scale=1]{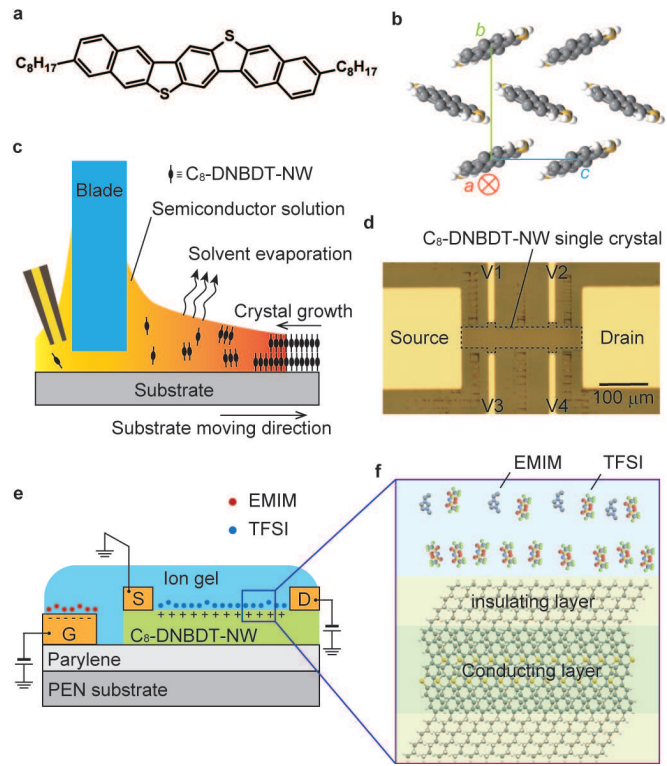}
		\caption{\textbf{Formation of electric double layer at the single-crystalline bilayer of \tm.} (a) Chemical structure and (b) crystal structure of \tm. The $c^*$-axis of \tm~corresponds to the carrier transport direction, where the effective mass $m^*$ is determined to be 1.1$m_0$~along the $c$-axis and 2.07$m_0$~along the $b$-axis\cite{Yamamura_SciAdv_2018}. (c) Schematic of the continuous edge-casting method. (d) Optical microscopy image of sample 2, in which the channel length and width are designed to be 250 $\mu$m and 60 $\mu$m, respectively. (e) Schematic of the present \tm\ EDLT. S, source electrode; D, drain electrode; G, gate electrode. (f) Schematic illustration of the ionic liquid/\tm\ interface.
		} 
		\label{fig1}
	\end{figure}

     As an ideal organic two-dimensional system, we employ an alkyl-substituted small molecule. A truly single-crystalline form of \tmfull \ (\tm, \ Figs.~\ref{fig1}a and b) is deposited on a flexible substrate via the continuous edge-casting method~\cite{Soeda_APEX_2013,Yamamura_SciAdv_2018,Kumagai_SciRep_2019} (Fig.~\ref{fig1}c). The continuous edge-casing method allows a one-shot crystal growth of layer-controlled thin films of \tm~with an areal coverage of up to at least 1 cm$^2$ ~\cite{Yamamura_SciAdv_2018,Kumagai_SciRep_2019}. This is large enough to cover an entire channel (Fig.~\ref{fig1}d). Technically, this one-shot printing method allows the fabrication of defect-free, single-crystalline nanosheets with large areal-coverage of up to 100 cm$^2$~\cite{Kumagai_SciRep_2019} onto any given substrate~\cite{makita2020high}. The single crystallinity across the \tm~thin film has been confirmed by X-ray diffraction and electron diffraction measurements~\cite{Yamamura_AdvFun_2020,makita2020high}. A uniform optical intensity in polarized microscopy images confirms the existence of a molecularly flat surface without molecular steps. A single-crystalline \textit{bilayer} is prepared selectively by controlling the substrate temperature~\cite{Yamamura_SciAdv_2018,Yamamura_CommunPhys_2020}. A plastic substrate (polyethylene naphtalate: PEN) coated with a parylene layer is a better alternative with regard to the thermal expansion coefficient of OSCs, $i.e.,$ less crystal cracks are induced during low temperature measurements. Gold/chromium electrodes are deposited gently on the surface of the bilayer \tm~to form source and drain electrodes (a photograph of the device is shown in Fig.~\ref{fig1}d). The channel direction is along the $c$-axis of \tm~. In this work, we use an ion gel as an electrolyte, in which the cations and anions of an ionic liquid (IL; 1-ethyl-3-methylimidazolium: [EMIM] and bis(trifluoromethylsulfonyl)imide: [TFSI], respectively) are enclosed by poly(vinylidene fluoride-\textit{co}-hexafluopropylene)(P(VDF-HFP)) networks. An ion gel sheet comprising [EMIM][TFSI] is then placed onto the channel and pre-deposited gate electrode to form a side-gate configuration (Fig.~\ref{fig1}e, details in Method section).\par
      Previous studies have revealed that the EDL formed particularly at the surface of the OSC causes an unavoidable dissolution/damage~\cite{Fukui_ChemCommun_2013}. In contrast, an insulating alkyl side chain of \tm~with a thickness of 1 nm can separate the conducting layer of \tm~ from the IL (Fig.~\ref{fig1}f). It is possible that the alkyl side chains provide indirect contact of the IL with the \tm~ molecules and suppress the electrical potential fluctuation, which is an origin of the carrier localization\cite{disorder_NatCommun_2016}. \par

		%%%%%%%%%%%%%%%%FIGURE 1
	%%%%%%%%%%%%%%%%%%

\begin{figure}[htbp]
		\centering
		\includegraphics[scale=1]{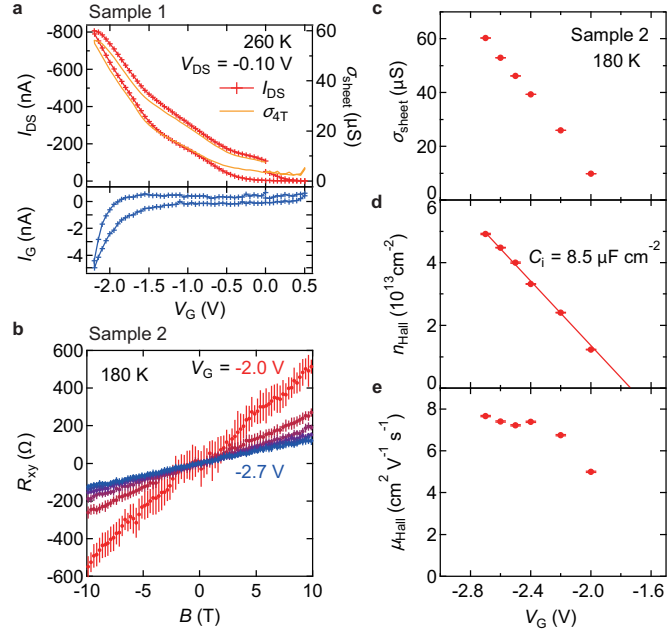}
		\caption{\textbf{High carrier density achieved with EDLTs.} (a) Transfer characteristics (\Id~vs \Vg) for sample 1 at $T$ = 260 K. \Id; drain current, \Ig; gate current, \Vd; drain voltage, \Vg; gate voltage. Transfer characteristics were acquired with a \Vg~sweep rate of 1.67 mV sec$^{-1}$. (b) Hall resistance \Rxy~profile with respect to an external magnetic field $B$ at various \Vg~at $T$ = 180 K for sample 2. \Vg~dependence of (c) sheet conductivity \sheetc~, (d) Hall carrier density \carrier~, and (e) Hall mobility \mob. The error bars in \sheetc, \carrier, and \mob~originate from the uncertainty in \Rxy and represent one standard deviation.}
		\label{fig2}
	\end{figure}

	To assess the formation of the EDL at the surface of \tm~, charging/discharging properties are investigated based on the standard EDLT characteristics. Figure~\ref{fig2}a shows the transfer characteristics of sample 1 at a temperature of $T$ = 260 K upon application of drain voltage \Vd~$= -0.10$ V. The drain current \Id~monotonically increases upon the application of negative gate voltage \Vg, showing a typical p-type operation. The gate current \Ig, which corresponds to the displacement current during charging/discharging, is found to be of the order of $10^{-9}$ A and two orders of magnitude smaller than \Id. This strongly suggests that charge carriers can be induced electrostatically rather than electrochemically. The formation of the EDL is further verified from the reproducible \Id~scans of \Vg. Four-terminal sheet conductivities \sheetc~$= \left( L_{\mathrm{4T}}/W\right) \left( I_{\mathrm{D}}/V_{\mathrm{4T}}\right) $~as a function of \Vg~($L_{\mathrm{4T}}$ are the lengths between the four-terminal voltage probes ($W$ is the channel width, and $V_{\mathrm{4T}}$ is the potential difference between the four-terminal voltage probes) and are found to reach up to $ca.$ 60 $\mu\mathrm{S}$ (equivalently, the sheet resistance of $\sim$ 17 k$\Omega$, see Fig.\ref{fig2}a for sample 1) at $T$ = 260 K and is at least five times larger than those attainable with conventional field effect transistor (FETs) with solid-state dielectrics~\cite{Yamamura_SciAdv_2018}. Of the four samples examined in this work, the highest \sheetc~was achieved for sample 1, and detailed \Vg- and $T$-variant \sheetc~was investigated for sample 2. We would emphasize that all the four samples show signs of metallic conduction and 2DHG behaviour at low temperatures (see the summary of sample variation in Extended data figure~\ref{figEX_EDLT}).\par
	
	The Hall effect for each sample was measured with a helium gas-exchanged cryostat. Figure~\ref{fig2}b shows the typical Hall resistance \Rxy~profile with respect to an external magnetic field $B$ for sample 2 at various \Vg~at $T$ = 180 K. The positive slope of \Rxy~with respect to $B$ indicates that the charge carrier accumulated at the interface of \tm~is hole, which is consistent with the p-type operation observed in the transfer characteristics. The sheet conductivity \sheetc~, Hall carrier density \carrier~, and Hall mobility at \mob~with various \Vg~, which are derived using the standard procedure of Hall effect measurements, are summarized in Figs.~\ref{fig2}c--e. \sheetc~and \carrier~increase linearly with increasing $|$\Vg~$|$. Linear fitting of \carrier(\Vg)reveals that the capacitance of EDL between IL and \tm~is is 8.5 $\mu$F cm$^{-2}$, which is consistent with the literature value~\cite{Iongel_AdvMater_2012}. These results lead to the conclusion that the carriers are induced electrostatically, $i.e.,$ the contribution of electrochemical doping is negligible. \mob~at $T$ =180 K increases from 5.0 \mobunit~at \Vg~= $-$2.0 V to 7.7 \mobunit~at \Vg~= $-$2.7 V (Fig. \ref{fig2}e), which is slightly smaller than those obtained for FETs with solid-state dielectrics~\cite{Yamamura_SciAdv_2018}. The decrease in mobility for the two-dimensional electron systems has been often observed, particularly when a high electric field (normal to two-dimensional sheet) is applied~\cite{Takagi_MOSFET1994}. It is possibly because not only ordinary phonon scattering but also other scattering such as scattering by ionized impurity may contribute to the carrier transport~\cite{lee2018direct,chaudhuri2019polarization,Davies_2DEG_book}. \par

		%%%%%%%%%%%%%%%%FIGURE 2
	
	%%%%%%%%%%%%%%%%%%
	
		%%%%%%%%%%%%%%%%FIGURE 3
\begin{figure}[htbp]
	\centering
	\includegraphics[scale=1]{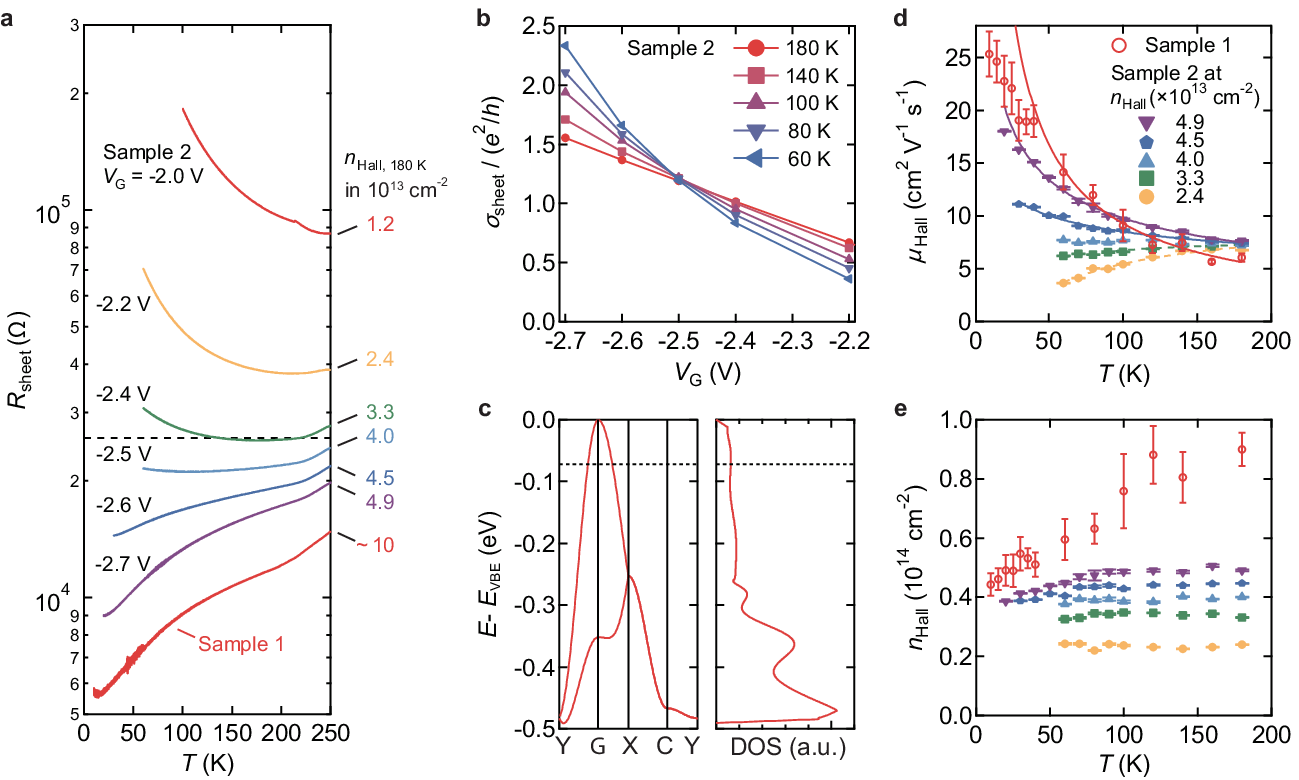}
	\caption{\textbf{Metal-insulator transition of \tm.} (a) Temperature $T$ dependence of the sheet resistance \sheetr~with at various \Vg. The dashed line represents the quantum resistance \QR. (b) \Vg~dependence of \sheetc~normalised with conductivity quantum $e^2/h$ for sample 2. (c) Energy band dispersion and integrated density-of-state (DOS) of valence band of \tm. The valence top ($E_{\mathrm{VBE}}$) positions at $\Gamma$-point. Here, $\Gamma = (0, 0, 0)$, X $= (0, \pi/b, 0)$, Y $= (0, 0, \pi/c)$ and C $= (0, \pi/b, \pi/c)$. The dashed line represents the position of Fermi energy when \carrier~$\sim$ 4.0 $\times$ 10$^{13}$ cm$^{-2}$ (critical point of metal-insulator transition). (d) $T$ Dependence of \mob~at various \carrier~($T = 180$ K). The solid curves represent the power law dependence $\mu \propto T^{-q}$, while the dashed curves represent the Arrhenius-type dependence of the thermal activation process \mob $\left( T\right) \propto$ exp$(-E_\mathrm{a}/(k_{\mathrm{B}}T))$. (e) $T$ dependence of Hall carrier density \carrier $=$ $(e R_{\mathrm{H}})^{-1}$ at various \Vg~. The error bars originate from the conductivity in \Rxy and represent one standard deviation.} 
	\label{fig3}
	
\end{figure}

%%%%%%%%%%%%%%%%%%

	Temperature-dependence of sheet resistance \sheetr \ $=$ \sheetc$^{-1}$ was investigated after the introduction of holes into the channel at $T$ = 260 K with various \Vg~(Fig.~\ref{fig3}a). At \Vg~= $-$2.0 to $-$2.4 V, a typical semiconducting behaviour is observed, in which the temperature-dependence changes from metallic behaviour due to carrier transport dominated by phonon scattering above 200 K to insulating behaviour below 200 K. The increase in $dR/dT$ with $\left|V_{\mathrm{G}}\right|$ indicates the gradual formation of degenerate holes, and enhanced mobility at low temperatures. Significantly, a positive temperature coefficient $dR/dT$ is observed when further holes are accumulated into the channel at \Vg~= $-$2.6 and $-$2.7 V, where a metallic signature $dR/dT > 0$ continues at the temperature of $T$ = 20 K. \sheetr(\Vg = $-$2.7 V)~is evaluated to be 20 k$\Omega$ at 260 K and decreases to 9 k$\Omega$ at 20 K for sample 2, and the minimum \sheetr~of 6 k$\Omega$ is achieved for sample 1. The positive $dR/dT$ is observed reproducibly for the other samples (see Extended data figure~\ref{figEX_RT}). Such an apparent signature of metal-insulator transition has never been achieved for heterointerfaces of OSCs. It can be clearly attributed to the strong carrier localization, which could not be overcome for any of the existing OSCs. materials~\cite{chwang2001temperature,takeya2005hall,podzorov2005hall,hulea2006tunable,takeya2007crystal,alves2008metallic,xia2010carrier,wang2012hopping,sakanoue2010band}. We emphasize that high-quality crystals even in the absence of surface crystal steps would be essential for the emergence of a 2DHG system at the heterointerfaces. Figure \ref{fig3}b shows the \sheetc~of sample 2 normalized by conductivity quantum $e^2/h$ as a function of \Vg. The temperature-independent crossing point can be seen at a well-defined \Vg~= $-$2.5 V (\carrier~= 4 $\times$ 10$^{13}$ cm$^{-1}$, equivalently 0.1 holes per molecule), which clearly segregates the metallic and insulating phases~\cite{Yamamoto2016MottIns}. The emergence of metallic ground state in highly doped \tm~is also consistent with the Mott-Ioffe-Regel criterion $k_{\mathrm{F}} l_{\mathrm{e}} > 1$ ($k_{\mathrm{F}}$ = $\sqrt{2 \pi n_{\mathrm{2D}}}$ is the Fermi wave vector and $l_{\mathrm{e}}$ = $\hbar k_{\mathrm{F}} \sigma / \left( e^2 n_{\mathrm{2D}}\right) $ is the mean free path of the charge carriers)~\cite{MIR_limit}, suggesting that the two-dimensional hole gas can be realized at the interface between single-crystal \tm~and IL. \par

	The observation of metallic states in \tm \ EDLTs can be associated with the strongly correlated two-dimensional (2D) hole gas. As carriers induced by EDLT are confined in two dimension, the Coulomb interaction between holes result in a large $r_{\mathrm{S}}$, which is the ratio of the hole-hole interaction $E_{\mathrm{C}}$ and Fermi energy $E_{\mathrm{F}}$:
	\[
	 r_{\mathrm{S}} = \frac{E_{\mathrm{C}}}{E_{\mathrm{F}}} = \frac{n_{\mathrm{v}} m^* e^2}{4 \pi \varepsilon \hbar^2 \sqrt{\pi n_{\mathrm{2D}}}} 
	\]
	where $n_{\mathrm{v}}$ is the degenerate valley, $m^*$ is the effective mass of carriers, $\varepsilon$ is the dielectric constant of the semiconductor, and $n_{\mathrm{2D}}$ is the carrier density. Generally, when $r_{\mathrm{S}}$ is larger than 1, the Coulomb interaction is not negligible and causes two-dimensional metal-insulator transition (2D MIT). Given the $n_{\mathrm{v}} = 1$, $m^* = 1.51 \ m_0$ ($m_0$ is the mass of free holes) as an average effective mass in $bc$-plane of \tm~, typical dielectric constant of OSCs\cite{OSC_dielectric_2016} $\varepsilon = 3$ and carrier density obtained in this work $n_{\mathrm{2D}} = 4 \times 10^{13}$ cm$^{-2}$, we evaluate $r_{\mathrm{S}}$ to be $\sim 8.5$, which is comparable to those obtained for inorganic 2D systems (for example, $r_{\mathrm{S}} \sim$ 8 for Si metal-oxide-semiconductor FETs\cite{rs-SiFET_PRL_1999}, $r_{\mathrm{S}} \sim$ 5 for GaAs/AlGaAs heterostructure\cite{rs-GaAs_2002}, and $r_{\mathrm{S}} \sim$ 4 for monolayer MoS$_2$~\cite{MoS2_NatMater_2013}.).  This implies that highly doped \tm~can be classified into the strongly correlated 2D system. The strong Coulomb interaction is further evidenced by the determination of Fermi energy. According to a previous report~\cite{Sugawara_CommunPhys_2018}, the trap density of \tm~single crystal is sufficiently low, $i.e.$, the pinning of the Fermi energy in the band gap is negligibly small, so that the Fermi energy can be shifted effectively as the carrier density varies. Given typical values of \carrier~4 $\times$ 10$^{13}$ cm$^{-2}$, the Fermi energy is shifted by 71.8 meV below the top of the valence band at $T$ = 260 K, leading to the Fermi degeneracy of \tm~ (Fig.~\ref{fig3}c). \par
	
	Figures~\ref{fig3}d and e display the $T$-dependence of \mob~and \carrier~ at various \Vg~of samples 1 and 2 (see Extended data figure~\ref{figEX_HE} for the other samples). \mob~measured below \carrier~$\sim$ 4.0 $\times$ 10$^{13}$ cm$^{-2}$ decreases as $T$ decreases, indicating that holes below the critical point undergo multiple-trapping-and-release conduction between the valence band and shallow trap states (within a few \kbT\ regions from the top of the valence band)~\cite{hasegawa2009organic}. We determine the activation energy $E_{\mathrm{a}}$ = 2.2 meV at \carrier~= 2.4 $\times$ 10$^{13}$ cm$^{-2}$ and 0.58 meV at \carrier~= 3.3 $\times$ 10$^{13}$ cm$^{-2}$ by fitting temperature-variant \mob~with the Arrhenius-type thermally activated model \mob $\left( T\right) \propto$ exp$(-E_{\mathrm{a}}/(k_{\mathrm{B}}T))$, which is small enough to activate from shallow trap states to valence band and is consistent with the weak temperature dependence of Hall carrier density \carrier~at low temperatures (Fig.~\ref{fig3}b). In contrast, \mob~measured above \carrier $\sim$ 4.0 $\times$ 10$^{13}$ cm$^{-2}$ increases monotonically as $T$ decreases (Fig.~\ref{fig3}b), where \mob~is estimated to be 7 \mobunit~at 180 K, and increases up to 26 \mobunit~at 10 K. The temperature-dependence of \mob~can be fitted with the power law behaviour $\mu \propto T^{-q}$ $\left( q > 0\right) $. The exponent $q$ in the power-dependence is estimated to be $q$ = 0.24 at \carrier~$\sim$ 4.5 $\times$ 10$^{13}$ cm$^{-2}$, $q$ = 0.46 at \carrier~$\sim$ 4.9 $\times$ 10$^{13}$ cm$^{-2}$, and $q$ = 0.86 at \carrier~$\sim$ 10 $\times$ 10$^{13}$ cm$^{-2}$. This is in good agreement with the exponent values of \mob~dominated by phonon scattering, which has been observed experimentally and theoretically for \tm~\cite{Tsurumi_NatPhys_2017}. Although the high \mob~exceeding 20 \mobunit~is reproducibly observed, a discrepancy in the temperature-dependence of \mob~from the power-law behaviour is found below $T$ = 50 K. This behaviour is consistent with typical 2D electron/hole systems, in which carrier transport is dominated by phonon scattering at elevated temperatures. Consequently, it is dominated by the scattering of defect impurities at cryogenic temperatures, resulting in suppressed mobility~\cite{Davies_2DEG_book}. In addition, the observation of positive longitudinal mangetoresistance is consistent with that expected in degenerated electron/hole systems. (see Extended data figure~\ref{figEX_MR}). \par

    \carrier~is estimated to be 1 $\times$ 10$^{14}$ cm$^{-2}$ at 200 K and 4 $\times$ 10$^{13}$ cm$^{-2}$ at 15 K. Remarkably high carrier density approaching 1 $\times$ 10$^{14}$ cm$^{-2}$ at room temperature corresponds to 0.25 holes per molecule, which is the highest carrier density observed in OSCs to the best of our knowledge, and clearly shifts the Fermi energy below the top of the valence band, leading the Fermi degeneracy. The decrease in \carrier~as temperature decreases does not imply a decrease in the actual carrier density because EDL can electrostatically confine the net of holes at the interface. In fact, this trend has been often observed for many of 2D electron/hole gasses and is interpreted as the crossover in different carrier scattering mechanism, most likely the cross over from phonon scattering at higher temperature to ionized impurity scattering at lower temperatures. This interpretation does not contradict to the mobility saturation at lower temperatures (Fig.~\ref{fig3}d). Alternatively, it is envisaged that the strong hole-hole interaction may induce a soft gap at the Fermi energy, resulting in a reduction of density-of-states. Although further researches will be necessary to fully understand electronic properties in 2DHG in OSCs interface, the presence of degenerated hole systems concomitantly with the signature of metal-insulator transition has been demonstrated for the first time in the OSC heterointerface.  \par

		 This study unambiguously demonstrates a metal--insulator transition in OSCs at ambient pressure for the first time. The metallic gas state with high a carrier density of 0.25 holes per molecule and remarkably low sheet resistance of 6 k$\Omega$ is experimentally observed. In a striking contrast with the localized nature of electronic states in OSCs, the observation of 2DHGs manifests itself, such that the static disorder inevitably found in OSCs is overcome. One-shot printing of semiconductor ink at room temperature allows for the ideal production of self-assembled, two-dimensional molecular nanosheets with a large areal coverage, up to 100 cm$^2$. The technique employed in this study to achieve remarkably high carrier density can act as framework for the exploration of electronic phase transitions in strongly correlated electronic systems in OSCs.\par
	
\noindent
\textbf{Methods}\\
\textbf{Sample preparation}
	
	The electric double-layer transistors (EDLTs) studied in this work were fabricated on a polyethylene naphthalate (PEN) substrate. The PEN film (Teijin Ltd., Q65HA, 125 $\mu$m) was pre-baked at 150 \(^\circ\)C for 3 h to reduce internal stress and cooled down slowly to room temperature. After cleaning the PEN film with acetone and 2-propanol in an ultrasonic bath, parylene (dix-SR, KISCO Ltd.) was coated onto the PEN film via chemical vapor deposition, achieving a thickness of 200 nm. The single-crystal bilayer \tm~was grown on the parylene layer from 3-chlorothiophene with a concentration of 0.020 wt\% at 70 \(^\circ\)C via the continuous edge casting method~\cite{Soeda_APEX_2013}. Gold (62 nm)/chromium (8 nm) were evaporated on the parylene/PEN substrate and single-crystal \tm~through shadow masks to form the source, drain, side gate electrodes, and voltage probes. Laser etching (V-Technology Co., Ltd., Callisto (266 nm)) was carried out for the \tm~layer to form Hall bar. \par
	The ion gel film was spin-coated onto the PEN film (Teijin Ltd., Q51, 25 $\mu$m) with a solution composed of poly(vinylidene fluoride-\textit{co}-hexafluopropylene)(P(VDF-HFP)), ionic liquid (1-Ethyl-3-methylimidazolium: [EMIM] and bis(trifluoromethylsulfonyl)imide: [TFSI], respectively) and, acetonitrile (1:2:10 wt. ratio). After cutting the ion gel with the PEN film, EDLTs were completed by the lamination of the ion gel and the PEN film onto the Hall bar-shaped single-crystal \tm~and side-top gate. \par 
	
 \textbf{Transport measurement}

    Low-temperature magnetotransport measurements were performed at least 12 h after the EDLT sample was inserted into a He gas-exchange cryostat with a superconducting magnet. Transfer characteristics in the linear regime (at drain voltage \Vd~$= -0.10$ V) were continuously recorded by a sweeping gate voltage \Vg~at a rate of $50$ mV per $30$ s. Introduction of the holes onto \tm~at 260 K was undertaken by applying \Vg~, for which \Vg~was negatively increased from $0$ V to the target voltage at a rate of $50$ mV per $30$ s. When \Vg~was changed, samples were heated up to 260 K and discharged with \Vg~$= 0$ V. After charging the \tm~by applying the target \Vg~at 260 K, the temperature $T$ dependence of the sheet resistance \sheetr~is measured by monitoring the drain current \Id~and the voltage probes. Constant \Vd~$= -0.10$ V is applied during the slow ramping of $T$ downwards and upwards with the ramping rate of $0.2$ K min$^{-1}$. Hall measurements were performed with a constant DC \Id~($1 \mu$A. During sweeping, an external magnetic field $B$ was applied perpendicular to the sample plane, not exceeding $\pm 10$ T and with a slow sweeping rate of 0.01 T s$^{-1}$.	
    
\textbf{Determination of Fermi level}

    The observation of metallic states in single-crystal \tm~with a high carrier density exceeding $4 \times 10^{13}$ cm$^{-2}$ in this work is accompanied by Fermi degeneracy. To investigate the relationship between the carrier density $n$ and Fermi level \CP~, we estimate the Fermi level of \tm~from the following equation:	 
    \begin{equation}
    n = \int_{-\infty}^{\infty} D(E) \left( 1 - f(E) \right) dE
    \end{equation}
    Where $D(E)$ is the density of states of \tm~and $f(E)$ is the Fermi distribution function $f(E)=[1+\mathrm{\exp}((E-$\CP$)/($\kbT$))]^{-1}$. The carrier density $n$ is estimated from the Hall effect at 180 K, and is assumed to be temperature-independent at least within the temperature from 180 K to 260 K. The density of states of \tm~is calculated using CRYSTAL 17 at the B3LYP functional and POB-DZVP basis set level.

%%%%%%%%
	\bigskip\noindent
	\textbf{Acknowledgements}\\
S.W. acknowledges the support from the Leading Initiative for Excellent Young Researchers of JSPS. T.O. acknowledges the support from PRESTO-JST through the project "Scientific Innovation for Energy Harvesting Technology" (Grant No. JPMJPR17R2). This work was supported by Kakenhi Grants-in-Aid (Nos. JP17H06123, JP17H06200, 20H00387) from JSPS. \\
%%%%%%%%%%%%%%	
	
	\bigskip\noindent
	\textbf{Author contributions}\\
N.K. conceived, designed, and performed the experiments and analysed the data. N.K. and J. Tsurumi performed the DFT calculations. T.O. synthesized and purified \tm. N.K., S.W. and J. Takeya wrote the manuscript.  S.W. and J. Takeya supervised the work. All authors discussed the results and reviewed the manuscript.

%%%%%%%%%%%%	
	\bigskip\noindent
	\textbf{Competing financial interests}\\
The authors declare no competing interests.

%%%%%%%%%%%%	
	\bigskip\noindent
	\textbf{Data availability}\\
The data that support the plots within this paper and the other findings of this study are available from the corresponding author (Shun Watanabe; swatanabe@edu.k.u-tokyo.ac.jp) upon request.

	\subsection{Reference}
	%\bibliography{Kasuya_MIT_library.bib}

\clearpage
    
%Extended Data Figures\\

\renewcommand{\figurename}{\textbf{Extended Data Figure}}
\setcounter{figure}{0}
%%%%%%EX1%%%%%%
\begin{figure}
\centering
\includegraphics[scale=1]{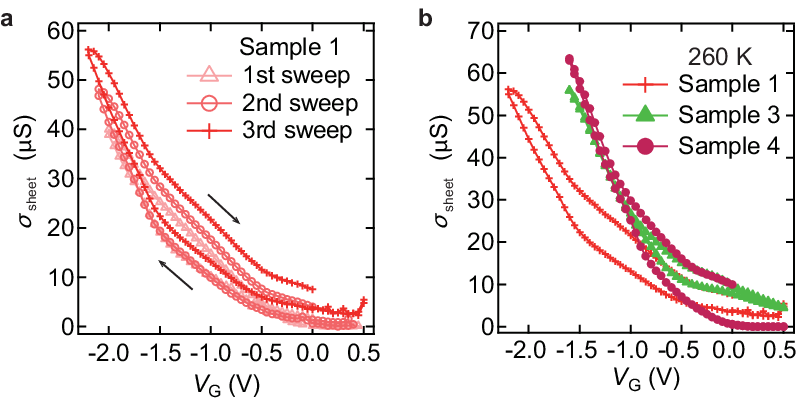}
\caption{\textbf{\Vg~dependence of \sheetc~at 260 K.} The reproducibility of multiple gate sweep (a) and the sample-dependence (b) of sheet conductivity \sheetc~at 260 K, respectively, obtained by \sheetc~$= \left( L_{\mathrm{4T}}/W\right) \left( I_{\mathrm{D}}/V_{\mathrm{4T}}\right) $~as a function of \Vg~($L_{\mathrm{4T}}$ is the length between the four-terminal voltage probes, $W$ is the channel width, and \Vxx~ is the potential difference between the four-terminal voltage probes). All measurements are performed with \Vd~= $-$0.10 V during \Vg~sweeping at a rate of 50 mV per 30 s. \sheetc~reaches around 60 $\mu$S in all samples reproducibly, although threshold voltage depends on the samples.}
\label{figEX_EDLT}
\end{figure}
\clearpage

%%%%%%EX2%%%%%%

\begin{figure}[htbp]
	\centering
\includegraphics[scale=1]{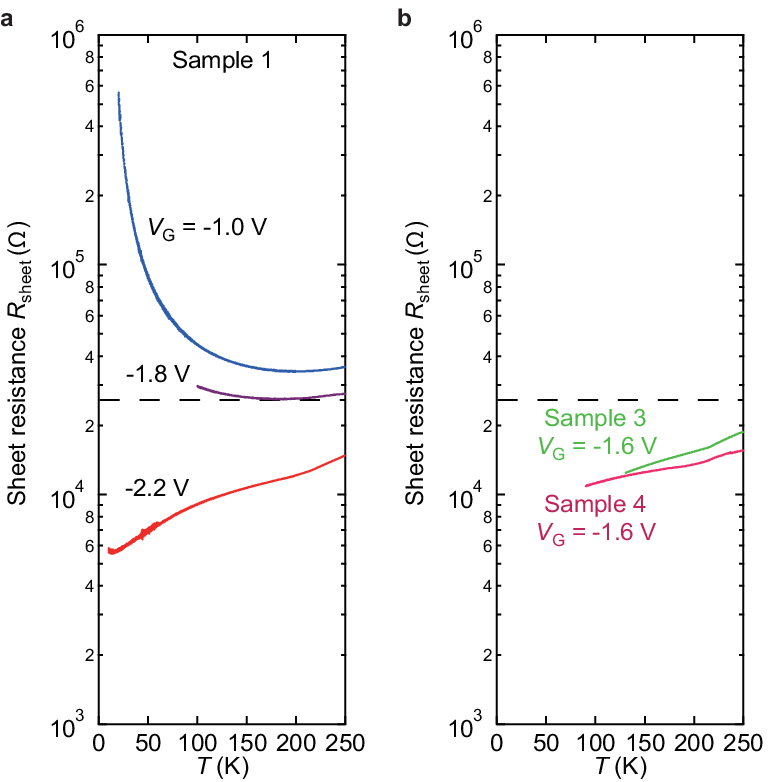}
	\caption{\textbf{The temperature-dependence of sheet resistance \sheetr.} The temperature $T$ dependence of sheet resistance \sheetr~($=$ \sheetc$^{-1}$) of sample 1 (a), 3 and 4 (b) (the \sheetr($T$) of sample 2 is described in the main text). In sample 1, insulator-to-metal crossover around the quantum resistance $h/e^2$ (black dash line in the panel) is observed with negatively increasing \Vg~from $-$1.0 V to $-$2.2 V. Positive $dR/dT$ is observed reproducibly for the samples 3 and 4.} 
	\label{figEX_RT}
\end{figure}
%\clearpage

%%%%%%EX3%%%%%%

\begin{figure}[htbp]
	\centering
\includegraphics[scale=1]{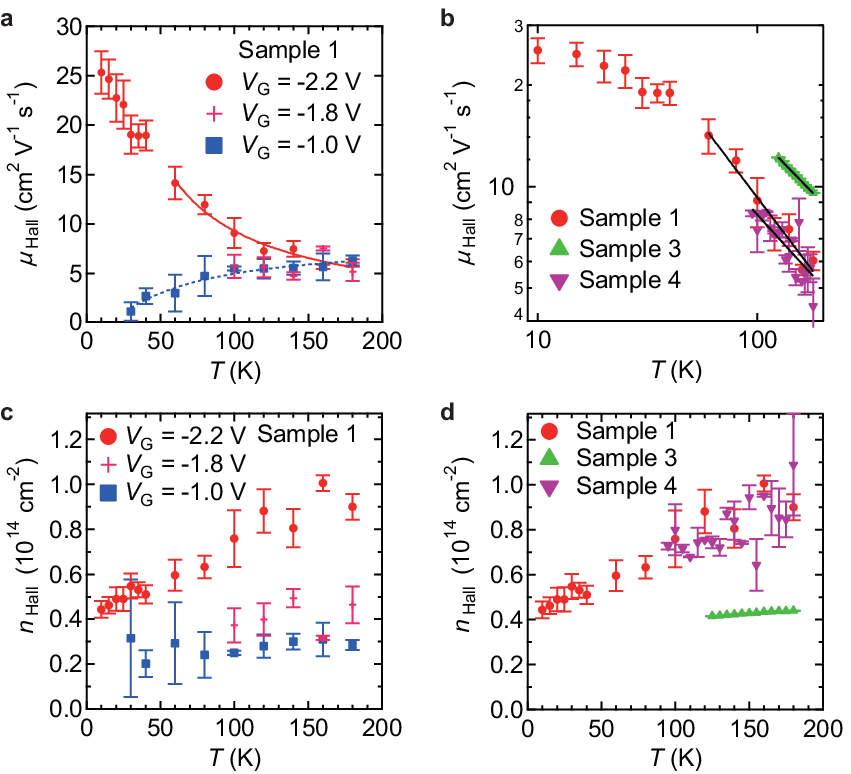}
	\caption{\textbf{The temperature-dependence of Hall mobility \mob~and Hall carrier density \carrier.} (a) and (c) The temperature-dependence of \mob~and \carrier~respectively of sample 1 at various \Vg. (b) and (d) The temperature-dependence of \mob~and \carrier~respectively of sample 3 at \Vg $= -1.6$ V and sample 4 at \Vg $= -1.6$ V compared with that of sample 1 at \Vg~= $-$2.2 V. For sample 1, \mob~at \Vg~ = $-$1.0 V decreases as temperature decreases with the activation energy $E_{\mathrm{a}}$ = 4.2 meV obtained by fitting with an Arrhenius-type thermally activated model \mob $\left( T\right) \propto$ exp$(-E_{\mathrm{a}}/(k_{\mathrm{B}}T))$ (see the dotted curve in (a)). By contrast, the temperature-dependence of \mob~at \Vg~= $-$2.2 V clearly increases monotonically with a decrease in temperature (a), where \mob~is estimated to be 5 \mobunit~at 200 K, and increases up to 25 \mobunit~ at 15 K. This typical metallic temperature behaviour is also observed in sample 3 at \Vg~= $-$1.6 V and sample 4 at \Vg~= $-$1.6 V in the right-top panel (b). (c) The Hall carrier density \carrier~is estimated to be 1 $\times$ 10$^{14}$ cm$^{-2}$ at 200 K and 4 $\times$ 10$^{13}$ cm$^{-2}$ at 15 K. (d) The temperature-dependence of \carrier~is also observed in sample 4 with \carrier $\left(180 \mathrm{K}\right) $ = 1 $\times 10^{14}$ cm$^{-2}$.} 
	\label{figEX_HE}
\end{figure}
\clearpage

%%%%%%EX4%%%%%%

\begin{figure}[htbp]
	\centering
\includegraphics[scale=1]{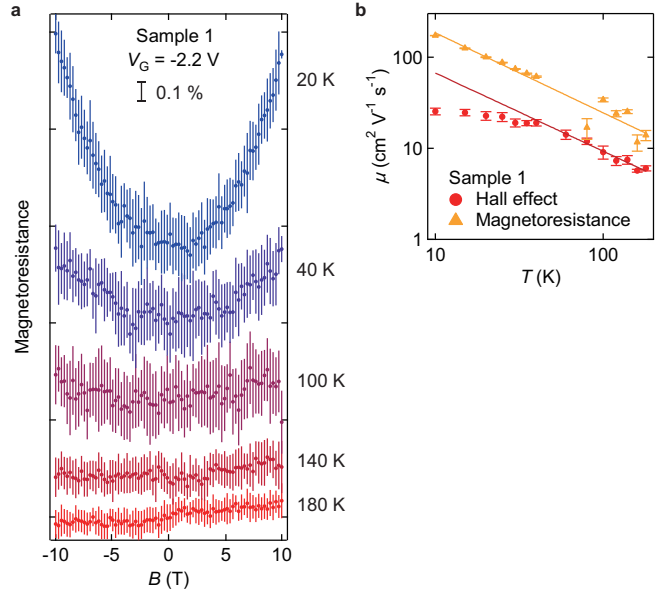}
	\caption{\textbf{The magnetoresistance of sample 1 at \Vg~= $-$2.2 V.} (a) The longitudinal magnetoresistance of sample 1 at \Vg~= $-$2.2 V defined as $(R (B) -R (0))/R (0)$ with respect to the application of $B$ perpendicular to the sample plane. The positive magnetoresistance is observed over a wide temperature range, which is in a good agreement with the well-established Lorentz magnetoresistance. In weak magnetic fields, the magnetotransport can be described via the semiclassical Boltzmann transport framework; the positive magnetoresistance is expected with a parabolic dependence to the applied magnetic field, $i.e.,$ $(R (B) -R (0))/R (0) = {\mu_{\mathrm{MR}}}^{2}B^{2}$. Here, the mobility $\mu_{\mathrm{MR}}$ is the only fitting parameter that can reproduce the magnitude of the positive magnetoresistance (shown in black curves). We summarize the temperature-dependence of the two mobilities determined from the Hall effect (\mob) and longitudinal magnetoresistance ($\mu_{\mathrm{MR}}$) in Fig. (b). While the temperature-dependence \mob~exhibits a saturation behaviour as temperature decreases, which is interpreted as a crossover in hole transports from phonon scattering to ionized impurity scattering, $\mu_{\mathrm{MR}}$ increases monotonically as temperature decreases with $\mu_{\mathrm{MR}} \propto T^{-q}$ $\left( q = 0.87\right)$. This discrepancy can be explained by the difference in scattering mechanisms.} 
	\label{figEX_MR}
\end{figure}
%\clearpage

\end{document}